%% file: main.tex
\documentclass[journal]{template/vgtc}                     

\onlineid{1820}

\vgtccategory{Methodological paper}

\vgtcpapertype{application/design study}

\title{Towards the Automatic Detection of Vection in Virtual Reality Using EEG}
\author{Gaël Van der Lee\thanks{e-mail: gael.vanderlee@univ-lille.fr}\\
    \scriptsize University of Lille
\and Anatole Lécuyer\\
    \scriptsize Inria Rennes\thanks{anatole.lecuyer@inria.fr}
\and Maxence Naud\\
    \scriptsize University of Lille
\and Reinhold Scherer\thanks{e-mail:r.scherer@essex.ac.uk}\\
     \scriptsize University of Essex
\and François Cabestaing\thanks{e-mail:francois.cabestaing@univ-lille.fr}\\
    \scriptsize University of Lille
\and Hakim Si-Mohammed\thanks{e-mail:hakim.simohammed@univ-lille.fr}\\
    \scriptsize University of Lille}

\abstract{%
  \input{contents/abstract}
}


\graphicspath{{figs/}{figures/}{pictures/}{images/}{./}{template/}} 

\usepackage{tabu}                      
\usepackage{booktabs}                  
\usepackage{lipsum}                    
\usepackage{mwe}                       

\usepackage{mathptmx}                  

\usepackage{manfnt}

\PassOptionsToPackage{svgnames}{xcolor}
\usepackage{pgfplotstable}
\pgfplotsset{compat=1.18}
\usepackage{multirow}
\usepackage{multicol}
\usepackage{soul}

\newcommand{\hsm}[1]{{\color{blue}{[H: #1]}}}
\newcommand{\fc}[1]{{\color{red}{[F: #1]}}}
\newcommand{\gv}[1]{{\color{orange}{[G: #1]}}}
\newcommand{\rs}[1]{{\color{green}{[R: #1]}}}
\newcommand{\al}[1]{{\color{violet}{[A: #1]}}}

\renewcommand{\hsm}[1]{}
\renewcommand{\fc}[1]{}
\renewcommand{\gv}[1]{}
\renewcommand{\rs}[1]{}
\renewcommand{\al}[1]{}

\usepackage{todonotes}

\usepackage{svg}

\begin{document}

\firstsection{Introduction}

\maketitle

\input{contents/introduction}

\section{Related Work}\label{sec:related-works}
\input{contents/related-work}

\section{Experiment}\label{sec:methods}
\input{contents/material-and-methods}

\section{Results}\label{sec:results}
\input{contents/results}

\section{Discussion}\label{sec:discussion}
\input{contents/discussion}

\section{Conclusion}\label{sec:conclusion}
\input{contents/conclusion}


\bibliographystyle{template/abbrv-doi-hyperref}

\bibliography{references}

\end{document}

%% file: contents/abstract.tex
Vection, the visual illusion of self-motion, provides a strong marker of the VR user experience and plays an important role in both presence and cybersickness. 
Traditional measurements have been conducted using questionnaires, which exhibit inherent limitations due to their subjective nature and preventing real-time adjustments.
Detecting vection in real time would allow VR systems to adapt to users' needs, improving comfort and minimizing negative effects like motion sickness.
This paper investigates the presence of vection markers in electroencephalogram (EEG) brain signals using evoked potentials (brain responses to external stimulations).

We designed a VR experiment that induces vection using two conditions: (1) forward acceleration or (2) backward acceleration.
We recorded both electroencephalographic (EEG) signals and gathered subjective reports on thirty (30) participants.
We found an evoked potential of vection characterized by a positive peak around 600 ms (P600) after stimulus onset in the parietal region and a simultaneous negative peak in the frontal region.
Our results also found participant variability in sensitivity to vection and cybersickness and EEG markers of acceleration across subjects.
This result is promising for potential detection of vection using EEG and paves the way for future studies towards a better understanding of vection.
It also provides insights into the functional role of the visual system and its integration with the vestibular system during motion-perception.
It has the potential to help enhance VR user experience by qualifying users' perceived vection and adapting the VR environments accordingly.

%% file: contents/introduction.tex
In Virtual Reality, vection -- the visual illusion of self-motion induced in a stationary observer \cite{palmisanoFutureChallengesVection2015} -- plays a critical role in shaping user experience.
This phenomenon occurs when a person experiences a sensation of movement while actually remaining stationary. 
An everyday example of this phenomenon is when an individual mistakenly perceives their own motion as a train departs from an adjacent platform, or when watching a river flow from a bridge.
Vection as a phenomenon has interested the scientific community for over a century\cite{dichgansVisualVestibularInteractionEffects1978, hettingerVisuallyInducedMotion1992} and has furthered our understanding of our perception of motion.

More recently, vection has been studied for its role in simulated environments such as driving or flight simulators, video games, and VR \cite{pohlmannEffectMotionDirection2021, rieckeCompellingSelfMotionVirtual2011}. 
Many have highlighted the need for a deeper understanding of this phenomenon \cite{palmisanoFutureChallengesVection2015, bertiNeuropsychologicalApproachesVisuallyInduced2020}.
Studying vection is important for several reasons. 
It could provide a functional understanding of the brain, particularly in terms of sensory integration between the visual and vestibular systems, leading to a better understanding of sensory conflict resolution.
It also elucidates how the brain judges speed and direction.
Moreover, vection is linked to postural control, with vision having a stabilizing effect on postural imbalance.
Third, vection contributes to a better understanding of the neural processes underlying motion perception. 
Fourth, vection is crucial for the efficacy of simulation and training programs.
Research is investigating whether vection is helpful to improve simulation skill transfer to the real world \cite{keshavarzVectionLiesBrain2015}.
Lastly, vection impacts motion sickness, and understanding it can lead to better prevention and treatment strategies for this condition. \cite{dichgansVisualVestibularInteractionEffects1978, palmisanoFutureChallengesVection2015}.
Additionally, a positive correlation has been found between vection and presence \cite{rieckeSceneConsistencySpatial2005, keshavarzEffectVisualMotion2019}.
Presence refers to the psychological state of "being there" \cite{cummingsHowImmersiveEnough2016} and serves as a key indicator of user experience in virtual reality (VR). 
It is a direct response to the degree of immersion provided by the VR environment \cite{slaterNotePresenceTerminology2003}.
Therefore, the examination of vection encompasses a multitude of implications with various related fields.

Cybersickness, commonly experienced in VR, is similar to motion sickness and often co-occurs with vection, prompting joint studies of the two phenomena \cite{keshavarzVectionVisuallyInduced2015, pohlmannEffectMotionDirection2021, nooijVectionMainContributor2017, kooijmanVirtualRealityStudy2022}. 
Moreover, vection can exacerbate cybersickness \cite{teixeiraUnexpectedVectionExacerbates2022, pohlmannRelationshipVectionCybersickness2022}.
Understanding the relationship between vection and cybersickness is crucial for designing VR experiences that enhance the user experience.

However, the primary challenge in studying these phenomena lies in the lack of well-validated, objective measures capable of consistently identifying or characterizing the vection experience \cite{palmisanoFutureChallengesVection2015}.
Currently, researchers often rely on non-standardized \cite{keshavarzVectionLiesBrain2015} subjective rating scales, leading to substantial variability across subjects and studies \cite{palmisanoFutureChallengesVection2015, bertiEarlyCorticalProcessing2019}.
Recent endeavors have emphasized the need for objective measurements of vection, arguing that subjective questionnaires for measuring vection suffer from misreported onset latencies, difficulty in obtaining real-time measurements, susceptibility to experimenter influence, potential confusion with other sensations, and the need for confidence in capturing true self-motion perception \cite{palmisanoFutureChallengesVection2015}.
However, this does not mean that questionnaires lack a place in vection research, as they are currently the best method for assessing a user's perceived experience.
As such, they can complement objective data.
This emphasizes the need to identify alternative indicators to complement and validate traditional self-report measures.
Effectively measuring vection would lead to more accurate data on self-motion perception by reducing reliance on subjective reports, enabling the use of objective indicators like eye movements and EEG. 
This is important because vection plays a crucial role in tasks like navigation, spatial orientation, and improving the user experience of virtual environments and simulators\cite{palmisanoFutureChallengesVection2015}.

Brain-Computer Interfaces (BCIs) are widely employed in the scientific literature to record, interpret, and convert brain signals into system inputs. 
BCIs use various techniques to measure brain activity, including electroencephalography (EEG), which records electrical potentials generated by neuron populations.
EEG is particularly suited for VR applications due to its relative ease of use and the allowance for user movement compared to other BCI methods.
Different types of signals can be extracted and analyzed from EEG data.
In the time domain, specific activation patterns, known as neuromarkers, can be identified under certain conditions.
An \textit{evoked potential} (EP) is a neuromarker that occurs in response to an external stimulus \cite{chiappaEvokedPotentialsClinical1997}.
In the spectral domain, patterns vary depending on the frequency range, reflecting conventional brain rhythms.
These rhythms are typically categorized as Delta (1--4 Hz), Theta (4--7 Hz), Alpha (8--13 Hz), and Beta (13--30 Hz).
A recent trend in BCIs involves passive BCIs, which monitor brain activity and adjust the system in response to the user's mental state.
This approach has been applied in VR to assess aspects of user experience, such as presence \cite{savalleElectrophysiologicalMeasurementPresence2024a}.

To the best of our knowledge, no evoked potential of subjective vection has been reported in the literature.
Our objective was to identify potential neuromarkers of vection through EEG.
In pursuit of this objective, we conducted an experiment that exposed our participants to moving white spheres in VR, while recording EEG.
The participant either experienced a forward acceleration (with the spheres accelerating in the backwards direction) or a backwards acceleration (with the spheres accelerating in the forward direction).
We then asked participants to rate the intensity of their perceived vection and analyzed how their self-reports correlated with their brain signals.
Our analysis yields the following key contributions:
\begin{itemize}
    \item We uncover differences in subject susceptibility to vection and acceleration direction.
    \item We replicate literature results finding bespoke signals of acceleration and showing an effect of vection on alpha power. 
    \item We identify an evoked potential of vection.
    \item We find a link between vection and Simulator Sickness Questionnaire answers.
\end{itemize}

The remainder of the paper is structured as follows: \autoref{sec:related-works} highlights the related work on the measurement of vection in virtual reality. \autoref{sec:methods} describes our experimental setup and methods. \autoref{sec:results} presents the findings of the investigation into the results from the experiment. Finally, we discuss shortcomings and future works in \autoref{sec:discussion} before concluding in \autoref{sec:conclusion}.

%% file: contents/related-work.tex
Vection as a term was first coined by Helmholtz in 1896 \cite{cz.HandbuchPhysiologischenOptik1896} by observing rivers flow under a bridge.
He hypothesized that it was an illusion, a failure of our senses.
Eighty years later, science had learned a lot more about this phenomenon and trends had started to emerge  \cite{dichgansVisualVestibularInteractionEffects1978}.
Advances showed that vection is more than an illusion, as visually induced motion is often necessary for an accurate representation of our movements through space.
The vestibular system alone is only able to detect acceleration and cannot infer velocity.
Thus, it cannot distinguish absence of motion to constant velocity, or a backward acceleration from a deceleration.
Other senses are needed to supplement the vestibular system and inform the brain's perception of motion.
The visual system plays an important role in informing our sense of motion along with the vestibular system.
This implies that the vestibular and visual system converge in the same locus to integrate into a unique model of spacial orientation  \cite{dichgansVisualVestibularInteractionEffects1978}.
Thus, studying vection offers insights into how that integration occurs and how our brains understand self-motion.

It was not until the 1970s that studies began to explore it through a physiological lens, as evidenced by the pioneering work of Dichgans and Brandt \cite{dichgansVisualVestibularInteractionEffects1978}.
Over time, the field has grown both in the number of publications, and in the breadth of applicable scenarios, ranging from simulators to rehabilitation
It can be used to monitor presence, improve sensorimotor training and rehabilitation\cite{keshavarzVectionLiesBrain2015, keshavarzVectionVisuallyInduced2015, bertiNeuropsychologicalApproachesVisuallyInduced2020}.
A review by Berti and Keshavarz \cite{bertiNeuropsychologicalApproachesVisuallyInduced2020} highlights the relevance of neuropsychological vection research, outlining 4 major reasons: (1) it helps uncover the neuro-cognitive functioning of multisensory perception (2) it presents an opportunity in research in other research areas (3) it can help develop an objective measure of vection, complementing subjective assessments; and (4) it has potential in neurorehabilitation.
A study by Seno et al. demonstrated a significant positive correlation between the duration of exposure to optic flow stimuli and the perceived strength of vection. 
Vection magnitude systematically increases with longer exposure durations, supporting the influence of exposure duration on vection strength \cite{senoVectionEnhancedIncreased2018}.
Kim et al. found that adding simulated visual movement, regardless of whether it was synchronized with head movements or viewed while stationary, consistently increased the strength and perceived speed of self-motion.
This suggests that visual processing plays a dominant role in the perception of self-motion, particularly when low-frequency sensory stimuli are involved \cite{kimEffectsActivePassive2008}.
Furthermore, correlations between dizziness and vection duration have been found, as well as between general discomfort and sway \cite{pohlmannEffectMotionDirection2021}. 
Moreover, unexpected vection significantly exacerbates cybersickness during HMD-based virtual reality, suggesting that unanticipated sensations of self-motion are a key predictor of cybersickness \cite{teixeiraUnexpectedVectionExacerbates2022}.
In many studies, a pivotal question remains: how to accurately measure vection?
In the subsequent sections, we will look into publications relating to the two primary categories of vection assessment: subjective and objective measurements.

\subsection{Subjective measures}\label{subsec:subjective-measures}
Subjective self-reports of vection have traditionally played an important role in vection research \cite{kooijmanMeasuringVectionReview2023, palmisanoIdentifyingObjectiveEEG2016}. 
Vection is commonly defined in studies as the visual illusion of self-motion \cite{palmisanoFutureChallengesVection2015}.
Questionnaire methods typically include binary choice responses, where participants indicate whether they experience vection, and onset time determination, which records when vection is first perceived. 
Intensity rating scales ask participants to rate the strength of their vection experience, while magnitude estimation involves quantifying the perceived intensity relative to a reference. 
These methods have been used to assess vection's occurrence, timing, and strength in various contexts. For example, participants might note the onset of vection during a VR simulation or rate its intensity on a scale \cite{kooijmanMeasuringVectionReview2023}.
These questionnaires have been used in studies to investigate various aspects, including their correlation with simulator sickness \cite{hettingerVectionSimulatorSickness1990} and their role in controlling self-motion and navigation\cite{palmisanoFutureChallengesVection2015}.

However, these methods have limitations stemming from their reliance on self-reporting.
While valuable, subjective measures can be influenced by individual differences and may lack the precision required for detailed analysis. 
Vection presents a dilemma for researchers, in that it is a subjective experience that would benefit to be studied across subjects.
In 2015, Palmisano et al. \cite{palmisanoFutureChallengesVection2015} highlighted four major challenges in modern vection research, which urged the community to (1) address diverse definitions and their implications, (2) explore the functional roles of conscious vection experiences during self-motion, (3) enhance objective measures for vection, and (4) conduct further research on the neural basis of vection to better understand its neural underpinnings. Our paper aims to contribute to the third and fourth challenges identified by Palmisano et al. 
 
 Multiple additional papers have also highlighted the need for objective and real-time measures of vection for validating research and making cross-subject and cross-study comparisons more robust \cite{keshavarzVectionLiesBrain2015, kooijmanMeasuringVectionReview2023, palmisanoIdentifyingObjectiveEEG2016}, citing various shortcomings, such as their unreliability, response biases or social desirability \cite{kooijmanMeasuringVectionReview2023}.

\subsection{Objective measures}\label{subsec:objective-measures}
The need for objective measures of vection has been recognized early on, as they can reveal the neural underpinnings of vection and enhance the design of VR systems. 
Understanding these neurophysiological aspects is crucial for improving VR experiences, as it allows for more informed control of user experiences and interactions.

Notably, postural sway analysis has been successfully employed to estimate the impact of vection on postural balance \cite{nooijVectionMainContributor2017, pohlmannEffectMotionDirection2021}.
This method provides valuable insights into the physical consequences of vection.
Such methods can be useful for understanding vection parameters.
For example, Palmisano et al. (2014)\cite{palmisanoSpontaneousPosturalSway2014} found that individuals who rely more on vision for postural stability tend to experience stronger vection.
However, it is hard to distinguish visually induced postural sway from visually induced vection \cite{palmisanoFutureChallengesVection2015}. 
This is due to the fact that visually induced postural sway shares similarities with visually induced vection, but it can occur without vection.
Recently, Padmanaban et al. \cite{padmanabanMachineLearningApproachSickness2018} have initiated efforts towards predicting vection directly from stereoscopic video. 
They employed a convolutional neural network-based optical flow algorithm, to compute features that are directly linked to the user's vection experience.

While such approaches advance our ability to predict vection based on visual input, understanding the neurophysiological underpinnings remains crucial.
These approaches lay the groundwork for understanding the neural mechanisms underlying vection.
Berti and Keshavarz \cite{bertiNeuropsychologicalApproachesVisuallyInduced2020} review EEG and fMRI studies on visually induced vection, identifying key brain areas involved in vection processing.
fMRI studies emphasize the involvement of a network in the neo-cortex in vection processing. 
EEG studies highlight changes in alpha band activity as well as an error related potential in the occipital


Electroencephalography (EEG) has emerged as a promising tool for objective vection assessment \cite{keshavarzVectionLiesBrain2015, bertiNeuropsychologicalApproachesVisuallyInduced2020, bertiEarlyCorticalProcessing2019}.
EEG allows for the investigation of neural correlates during vection experiences.
By exploring how vection is processed in the brain, VR developers can better control the environments, reduce cybersickness, and predict overall user experience.
There are several ways to investigate vection using EEG.
Looking at EPs, Thilo et al. found a negative occipital response 70ms after optokinetic stimulation \cite{thiloPerceptionSelfMotionPeripheral2003}, but did not link this neuromarker to vection.
Some studies looking into brain rhythms during vection found that alpha suppression is correlated with vection \cite{harquelModulationAlphaWaves2020, bertiNeuropsychologicalApproachesVisuallyInduced2020}.
It is important to note that, to investigate robust markers of vection in EEG, one still needs to rely on subjective measures, as there is no other ground truth measurement.
Since vection is fundamentally a subjective experience, any objective measure of vection must be measured in conjunction with traditional self-measures reports \cite{palmisanoFutureChallengesVection2015}.
Moreover, current EEG methods of studying vection still suffer shortcomings that prevent from using them as a ground truth for vection detection.
Many do not generalize between subjects, and methods able to isolate lower frequencies such as those found in the literature can result in longer time windows.
In this study, we focus on evoked potentials, as they provide a high temporal resolution and may reveal characteristic brain responses that are consistent across subjects, offering a potential objective measure of vection.
Our paper proposes a path for objective measurement of vection using EEG by leveraging the established field of Evoked Potentials.
Our proposition utilizes acceleration as a stimulus to explore EPs specific to the phenomenon of vection.

%% file: contents/material-and-methods.tex
The objective of this study is to investigate Evoked Potentials markers of Vection in VR\@.
Thus, we designed a user study which aims to trigger vection in two types of trials. 
This protocol follows the same procedure as outlined in \cite{vanderleeEEGMarkersAcceleration2024}, with the addition of questionnaires for assessing vection and simulator sickness.

The subject either feels (1) a sudden forward acceleration \texttt{$FA_1$}, with the environment moving in the posterior direction, or (2) a sudden backward acceleration \texttt{$BA_1$}, with the environment moving in the anterior direction.
After each trial, the subject is asked to report vection in a questionnaire.

\subsection{Participants and ethics}\label{subsec:participants-and-ethics}
Thirty healthy participants with normal or corrected to normal vision took part in the experiment  (18 men, 12 women aged $\mu=26$, $min=18$, $max=56$, $\sigma=7.38$).
This study was approved by the University of Lille Ethics Committee and adheres to the principles outlined in the Declaration of Helsinki.
All data was anonymized, and participants provided written informed consent.
They were also explicitly informed of their right to withdraw from the experiment at any time without any repercussions.

\begin{figure}
    \centering
    \includegraphics[width=.8\columnwidth]{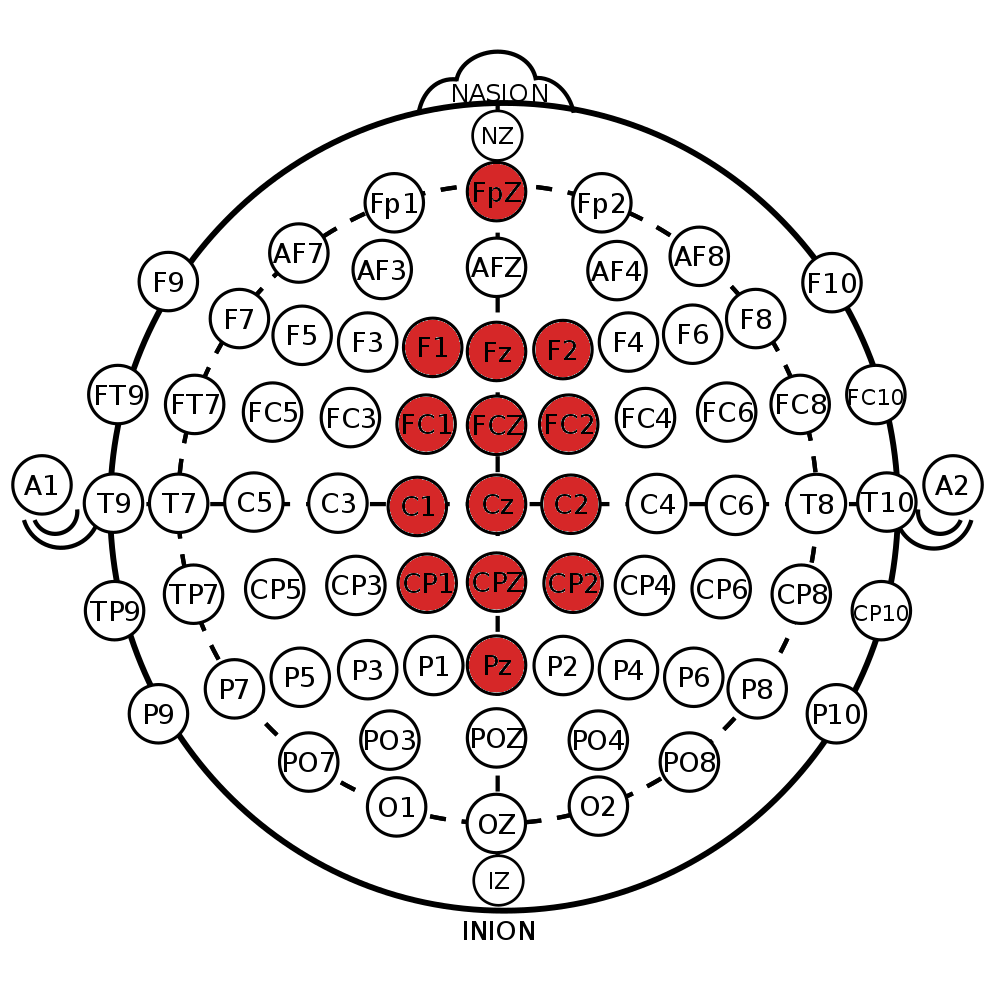}
    \caption{Electrode Placement in the High-Resolution 10–20 International System. The highlighted electrodes (marked in red) indicate the specific sites used in this study: FPz, Fz, F1, F2, FCz, FC1, FC2, Cz, C1, C2, CPz, CP1, CP2, and Pz.}
    \label{fig:electrode_placement}
\end{figure}

\subsection{Apparatus}\label{subsec:apparatus}
The virtual environment was displayed on a Valve Index HMD with a 1440x1600 resolution screen for each eye running at 144Hz and a DELL PRECISION 3640 personal computer with an NVIDIA GeForce RTX 3080 video card.

EEG was measured using a cap g.GAMMAcap2 from g.tec medical engineering GmbH®(Austria) with 14 recodring electrodes, plus a reference and ground electrode.
The recording electrode configuration was the following: FPz, Fz, F1, F2, FCz, FC1, FC2, Cz, C1, C2, CPz, CP1, CP2 and Pz.
The central distribution of the electrodes was purposefully chosen as we expected a non-lateralized signal along the frontal-occipital axis as seen in \autoref{fig:electrode_placement}.
The software used to record EEG data and events is the OpenVibe 3.1.0 software \cite{renardOpenViBEOpenSourceSoftware2010}.
The virtual environment (VE) was created using the version \textit{2020.3.11f1} of the Unity game engine software.
Data analysis was performed using the MNE-python library \cite{larsonMNEPython2023} and the seaborn visualization library \cite{l.waskomSeabornStatisticalData2021}.

\subsection{Trial design}\label{subsec:trial-design}

\begin{figure}[tb]
    \centering
    \includegraphics[width=\columnwidth]{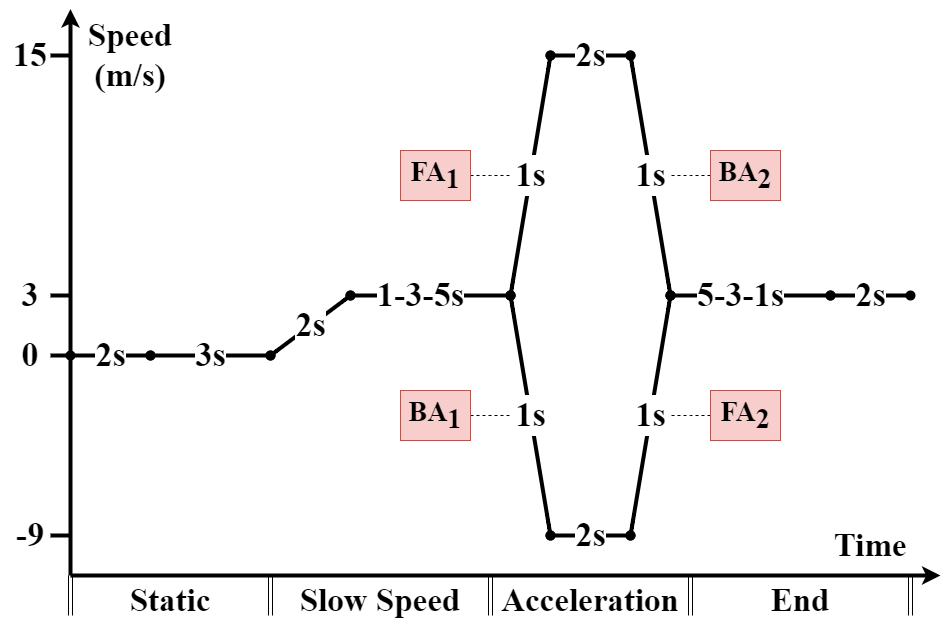} 
    \caption{Illustration of a trial: Depiction of the evolution of speed over time. Dashed lines represent variable delays, with one of the three delays chosen randomly. The second delay is selected to ensure that the cumulative sum of delays amounts to 6 seconds.}
    \label{fig:trial}
\end{figure}

A visual representation of a single trial is represented in \autoref{fig:trial}.
Each trial, lasting 19 seconds, was divided into four distinct phases.
First, during the \textit{Static phase}, the virtual environment gradually appeared over the course of 2 seconds and remained stationary.
In the \textit{Slow speed phase}, the environment accelerated to a speed of 3 m/s within 2 seconds and sustained this speed for a variable time of 1, 3 or 5 seconds, providing a baseline for EEG measurement under low-speed conditions.
The \textit{Acceleration phase} then followed, where participants experienced either a forward ($FA_1$) or backward acceleration ($BA_1$) of 12 m/s² for a duration of 1 second. 
These parameters were designed to ensure the acceleration feels sudden and the resulting speed appears fast while remaining within familiar, everyday ranges, such as those experienced in a car. 
Additionally, evoked potentials typically occur within 1 second of stimulus onset, and prolonged exposure to visual acceleration has been shown to produce diminishing perceptual effects \cite{goldsteinJudgmentsVisualVelocity1957}.
The speed was then held for 2 seconds before slowing back to the initial speed, either by forward ($FA_2$) or backward acceleration ($BA_2$).
Finally, in the \textit{End phase}, the environment continued moving at 3 m/s, mirroring the slow-speed phase, before gradually disappearing over a 2-second fade-out.

\subsection{Experimental setup}\label{subsec:experimental-setup}

For the design of the VE we could either induce vection in a realistic environment, which is more engaging for the participants, or we could do it in a minimalist environment.
The minimalist environment of white spheres arranged like a star field is more widely used in the literature \cite{bertiNeuropsychologicalApproachesVisuallyInduced2020}.
This is because it works well at inducing vection while keeping potential confounding factors at a minimum.
We chose to proceed with a minimalist environment.

\begin{figure}[tb]
    \centering
    \includegraphics[width=\columnwidth]{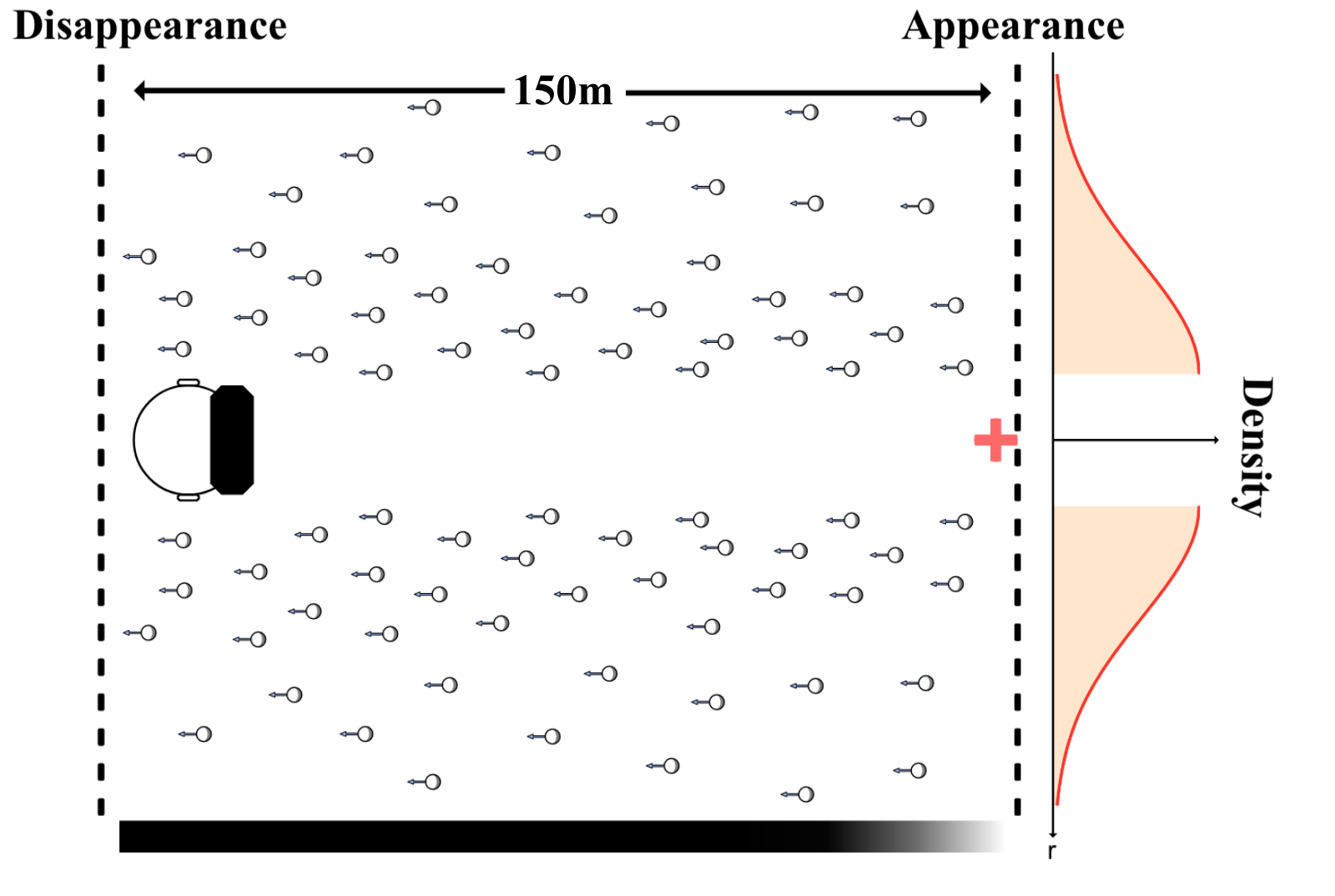}
    \caption{A top-down view of the simulation shows spheres uniformly distributed along the depth axis and arranged radially with a Gaussian distribution. Spheres become increasingly transparent as they approach the depth limit to prevent visual distraction. To avoid participant collisions, spheres are kept away from the radial center.}
    \label{fig:env}
\end{figure}

Our VE is similar to that used by Keshavarz et al. to induce vection \cite{keshavarzEffectVisualMotion2019}.
The experimental protocol involves seating the participant within a virtual environment surrounded by stationary white spheres. 
\autoref{fig:env} describes visual representation of sphere placement.
Initially, the participant remains static, but as the experiment progresses, they either accelerate forward or backward, serving as stimuli to induce vection.
This is realized using a 3D cloud composed of 2000 white spheres arranged cylindrically around the participant against a dark background in a virtual reality environment.
Dimensions in the VE are measured in ``units``, and calibrated such that the participant's height when standing is equivalent to 1.8 units.
This results in 1 unit being roughly equal to a meter.
Each sphere has a diameter of 0.20 meters.
The distribution is centered 1.2 meters above the ground to align with seated participants' eye level. 
They are uniformly distributed in terms of angle and depth within the range of 0 to 150 meters.

\begin{figure}[tb]
    \centering
    \includegraphics[width=\columnwidth]{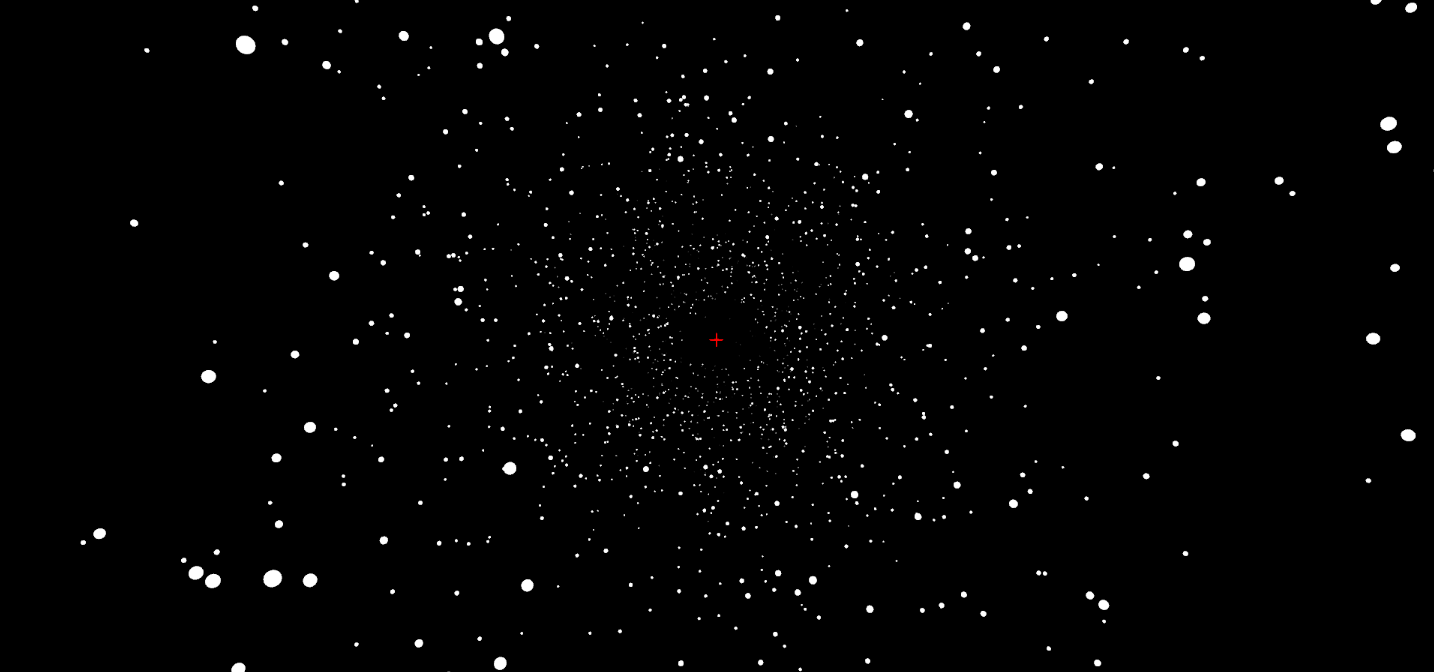}
    \caption{Depiction of the visual experience presented to participants, featuring a virtual scene composed of point clouds and a central crosshair.}
    \label{fig:stars}
\end{figure}

The participant's view can be seen in \autoref{fig:stars}.
A red crosshair is positioned at the center of the visual field, and participants are instructed to maintain their gaze fixed upon it to avoid ocular movements.
Participants are steated in a chair instructed to keep their head movements to a minimum during the experiment in order to minimze artifacts.
To achieve a homogenous distribution of the spheres, we employ a Gaussian distribution for the radial axis with a standard deviation of 5 meters.
To prevent spheres from coming too close to the participant's head, they are maintained 1.8 meters away from the radial center.
Spheres appear 150 meters away and gradually become visible over the first 30 meters to avoid distractions.
The movement of each sphere is synchronized and updated in real-time, facilitated by a separate computer handling acceleration components.

\subsection{Experimental procedure}\label{subsec:experimental-procedure}
Upon entering the experimental room, participants were briefed on the study's objective: detecting subjective sensations of motion patterns in EEG data.
We confirmed the absence of epilepsy or implanted electrical devices in participants.
Next, participants were briefed on the study's procedure and its various stages, including the devices, environment, stimulations, and the questions regarding vection.
The vection and vection scale were explained uniformly to all participants, as it is crucial to avoid describing either the participant or the spheres as moving, since both interpretations of the stimuli are valid and could bias their responses.

Then, informed consent was obtained, and each participant received a unique identification number and completed a demographic questionnaire, as well as a Simulator Sickness Questionnaire (SSQ) \cite{kennedySimulatorSicknessQuestionnaire1993}.
The SSQ is a 4-point Likert scale originally designed to assess simulator sickness in aviators.
It was selected due to its extensive use in cybersickness research, allowing for comparative analysis with other studies. 
It is a widely used tool that provides granular insights into participant well-being by categorizing responses into Oculomotor, Nausea, and Disorientation scores, thus enabling a focused assessment of symptoms related to cybersickness.
The relationship between an identification number and the subject's name was only known to the subject.
Sensors were then affixed to record EEG\@.

Subsequently, participants were seated and viewed 78 vection-inducing events organized into four blocks, with each block containing 20 trials, except for the fourth block, which consisted of 18 trials to keep all trials types balanced.
This means that the participant experiences 13 times each possible trial: Forward with 1, 3 or 5 second delay and Backward with 1, 3 or 5 second delay.

Following each block, participants were given a rest period of minimum 5 minutes to recover, during which they completed the SSQ.
The next bloc begins once the participant indicates his readiness.

Each trial comprised a 19-second visual simulation (see \autoref{subsec:trial-design}) followed by a brief period for oral rating. The participant was asked to rate the perceived vection intensity on a four-point Likert scale, staying consistent with the SSQ:
\begin{itemize}
    \itemsep0em
    \label{item:vection-scale}
    \item \textbf{NO VECTION}, which we will refer to as \textbf{\texttt{NV}}, indicating subject only perceived object-motion.
    \item \textbf{WEAK VECTION}, which we will refer to as \textbf{\texttt{WV}}, indicating subject perceived slight self-motion and mostly object-motion.
    \item \textbf{MODERATE VECTION}, which we will refer to as \textbf{\texttt{MV}}, indicating subject perceived self-motion and object-motion equally.
    \item \textbf{STRONG VECTION}, which we will refer to as \textbf{\texttt{SV}}, indicating subject only perceived self-motion.
\end{itemize}
Special care is taken to ensure that vection and the vection scale are explained consistently to all participants.

\subsection{Data Processing}\label{subsec:data-processing}

The processing pipeline utilizes MNE-python \cite{larsonMNEPython2023} for data handling and filtering.
First, channels exhibiting variance and noise abnormally higher than others were manually identified, marked as bad and excluded from the analysis.
Then the EEG data was re-referenced using common average referencing (CAR) and resampled to 128Hz.
Finally, it was filtered in the 0.3 to 10Hz range using a 4th order IIR forward and backward Butterworth filter for all plots except the power spectral density (PSD) plot.
Epochs range from 0.5s before stimulus to 1s after stimulus. 
Any epoch in which the EEG signals of a channel exceed 125$mV$ was rejected.

The data is stored using the EEG Brain Imaging Data Structure or BIDS format \cite{gorgolewskiBrainImagingData2016, pernetEEGBIDSExtensionBrain2019}, a standard for organizing and describing brain imaging datasets.
It allows researchers to readily organize and share study data within and between laboratories.
The data gathered for this study will be made publicly available.
Visualization was performed with the help of the seaborn library \cite{l.waskomSeabornStatisticalData2021}.

%% file: contents/results.tex
The data acquired in this study underwent analysis through two approaches. 
Initially, we examined which factors influenced subject ratings. 
We then conducted an aggregate analysis across all subjects, focusing on EEG patterns related to vection. 
Furthermore, we extend results obtained in the literature concerning responses to acceleration.

\subsection{Subjective Results}\label{subsec:behavioral-results}
We combined the subject's responses to our questionnaires and analyze the results.

\begin{figure}[tb]
    \centering
    \includegraphics[width=\columnwidth]{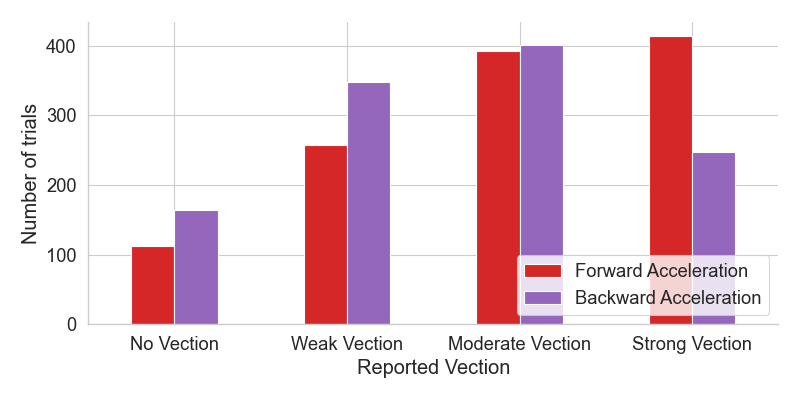}
    \caption{
    Distribution of reported vection per trial for $FA_1$ (red) and $BA_1$ (purple).
    }
    \label{fig:behavioral}
\end{figure}

The distribution of responses to the vection questionnaire can be found in \autoref{tab:reported_vection}.
The results show that some participants exhibited greater sensitivity to vection than others, reinforcing the need for VR systems that can adapt to individual user responses in real time to improve comfort and reduce cybersickness.
As seen in \autoref{fig:behavioral}, there are relatively fewer trials where a subject experienced \texttt{NV}.
Some subjects were more sensitive than others to the vection-inducing events.
A Kruskal-Wallis test revealed a significant difference in reported vection between subjects ($H=184.78$, $p<0.001$, $\epsilon^2=0.406$), emphasizing the variability in vection reporting across individuals.
Similar results are reported in the literature \cite{palmisanoFutureChallengesVection2015, bertiEarlyCorticalProcessing2019}.
\autoref{fig:behavioral} shows the distribution of reported vection for $FA_1$ and $BA_1$.
Chi-squared testing reveals that acceleration direction ($FA_1$ versus $BA_1$) has a strong influence on reported vection ($\chi^{2}=82.06$, $p<0.001$, $df=3$).

\begin{table}[tb]
    \centering
    \caption{
    Totals for reports of vection for each subject across all trials. Each row is a subject, the last row represents the average for each column.
    }
    \begin{filecontents*}{figures/reported_vection.csv}
Strong Vection,Moderate Vection,Weak Vection,No Vection
28,22,19,8
30,35,11,1
22,15,25,16
1,24,40,10
9,28,27,14
14,39,25,0
1,10,28,38
41,33,4,0
33,8,32,3
65,14,1,0
36,38,6,0
42,24,11,1
24,35,14,4
17,31,30,0
0,0,21,57
42,24,12,0
18,31,28,1
18,36,21,3
26,52,0,0
0,0,42,36
6,27,31,14
20,40,19,1
18,37,23,0
10,27,27,13
25,31,19,3
0,39,39,0
56,19,3,0
17,48,13,0
0,0,26,52
43,26,8,1
22,26,20,9
\end{filecontents*}
\pgfplotstabletypeset[
        every head row/.style={
            after row=\midrule,
        },
        col sep=comma,
        string type,
        header=has colnames,
        every last row/.style={before row=\bottomrule}
    ]{figures/reported_vection.csv}
    \label{tab:reported_vection}
\end{table}

Our objective is to examine the neural markers of vection and how they relate to subjective vection reports.
We also analyze the differences between conditions where vection was reported and those where it was absent.
The four-point vection scale provides valuable granularity; however, for the purpose of distinguishing the presence or absence of vection, a binary categorization is more appropriate.
The goal is to clearly separate trials where participants did not feel substantial self-motion from those where they reported strong self-motion.
This approach helps to draw a more definitive boundary between object-motion perception and vection, reducing ambiguity in the analysis.

The examination revealed variability in the perception threshold across subjects, with the \texttt{NV} category being notably underrepresented in the $FA_1$ condition, as evident in \autoref{fig:behavioral}. 
Given these considerations, we merged the \texttt{NV} and \texttt{WV} categories for comparative analysis against the \texttt{SV} category.
\texttt{MV} was excluded and served as a buffer to differentiate between the two classes.

\subsection{Correlates of Acceleration}\label{subsec:baseline-and-vection-events}

In our study, we divided the EEG data into different conditions. The baseline condition was recorded during the slow-speed phase, where the subject moved at a constant speed of 3 m/s, as shown in \autoref{fig:trial}. 
This ensures that some visual stimulation is also present in the baseline condition.
Additionally, the conditions labeled $FA_1$, $BA_1$, $FA_2$, and $BA_2$ were defined as described in the \textit{Trial Design} subsection.

To highlight the difference in EEG responses when vection occurs, we compare the median responses around accelerations and a baseline.
We selected the median over the mean due to its greater robustness against outliers and because some artifacts bypassed the artifact rejection process, potentially distorting the mean.
A Shapiro–Wilk test confirmed that the data is not normally distributed with a W statistic of $0.0247$ and $P<0.0001$.

We assess the significance of observed differences using a non-parametric bootstrapping method. We generate 10,000 resamples of our data with replacement, and then calculate 95\% confidence intervals, defined as the range between the 2.5th and 97.5th percentiles of the resampled data. These confidence intervals are displayed as shaded areas in the figures.

\begin{figure}[tb]
    \centering
    \includegraphics[width=\columnwidth]{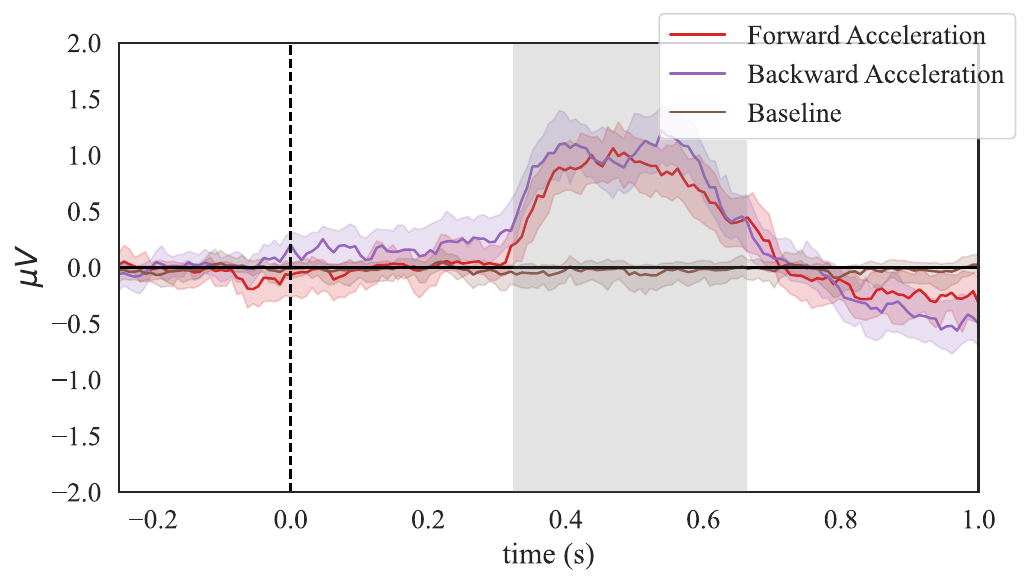}
    \caption{
    Comparison of the median of the FCz electrode for $FA_1$ (red), $BA_1$ (purple) and baseline (brown).
    The dotted line at 0 seconds represents the start of the acceleration.
    The y axis represents the median voltage for Cz at that time point.
    The 95\% confidence interval is displayed around each event's line, in its corresponding color.
    The shaded gray area corresponds to the time period where there was a statistically significant difference between the signals.
    $FA_1$ and $BA_1$ present similar patterns distinguishable from the baseline.
    }
    \label{fig:baseline}
\end{figure}

\begin{figure}[tb]
    \centering
    \includegraphics[width=\columnwidth]{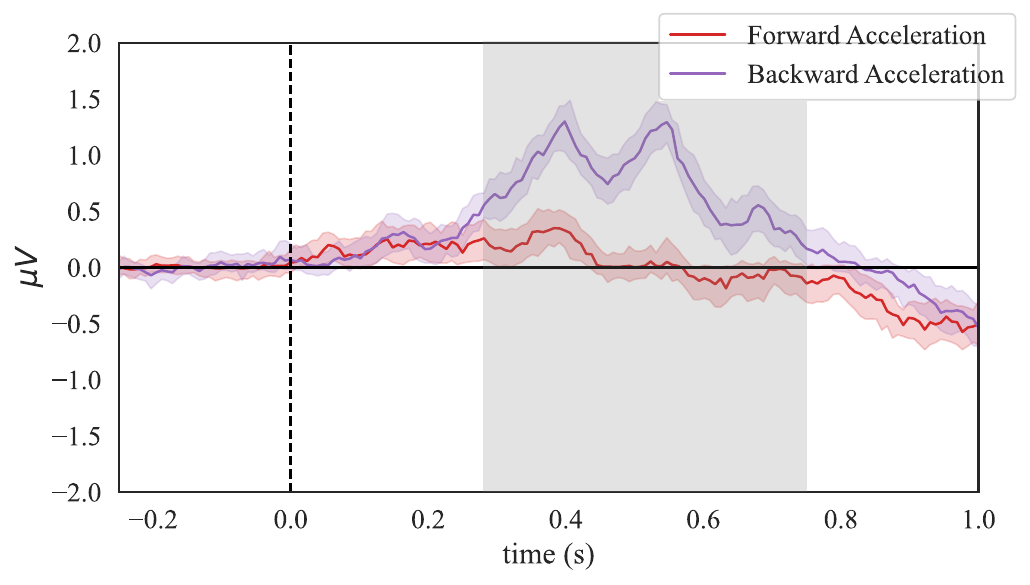}
    \caption{
    Comparison of the median of the Cz electrode for $FA_1$ (red), $BA_1$ (purple).
    The dotted line at 0 seconds represents the start of the acceleration.
    The y axis represents the median voltage for Cz at that time point.
    The 95\% confidence interval is displayed around each event's line, in its corresponding color.
    The shaded gray area corresponds to the time period where there was a statistically significant difference between the signals.
    The Cz presents a significant difference between $FA_1$ and $BA_1$.
    }
    \label{fig:acc_dec}
\end{figure}

Our findings on acceleration reinforce previous results reported in the literature \cite{vanderleeEEGMarkersAcceleration2024} by expanding the subject cohort and finding similar patterns of acceleration perception.
\autoref{fig:baseline} shows EEG signals that present statistically significant differences when the subject is experiencing $FA_1$ or $BA_1$ compared to baseline on the FCz electrode.
This signal is identical in both the $FA_1$ and $BA_1$ conditions, but it differs significantly from baseline.
It represents a marker of visual acceleration.
\autoref{fig:acc_dec} shows a significant difference between $FA_1$ and $BA_1$ on Cz.
This presents a characteristic signal of the acceleration direction.

\subsection{Correlates of vection}\label{subsec:correlates-of-subjective-vection}

\begin{figure}[tb]
    \centering
    \includegraphics[width=\columnwidth]{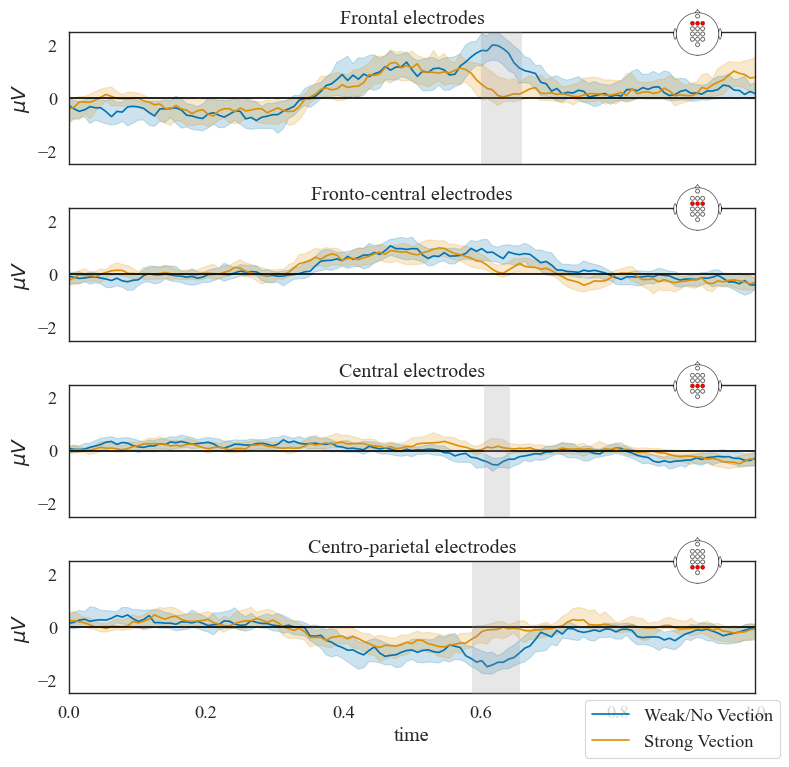}
    \caption{
    Comparison of the EEG signals between \texttt{WV} \& \texttt{NV} (blue) and \texttt{SV} (orange) during $FA_1$.
    The x axis represents the time since the acceleration started.
    The y axis represents the median voltage of the electrodes at that time point.
    The 95\% confidence interval is shown around the line of each vection category, in its corresponding color.
    The shaded gray area corresponds to the time period where there was a statistically significant difference between the signals.
    The \texttt{SV} signal has a significant positive deviation from baseline around 600ms after acceleration onset compared to a negative deviation for \texttt{WV} \& \texttt{NV}.
    This trends reverses as the signal travels from the parietal to the frontal region, with a negative deviation for \texttt{SV} at 600ms compared to a positive deviation for \texttt{WV} \& \texttt{NV}.
    This graph is based on 92 trials for the \texttt{WV} \& \texttt{NV} and 87 trials for the \texttt{SV}.
    }
    \label{fig:vection_forward}
\end{figure}

Addtionally, we find novel results concerning correlates of self-motion perception. 
The data shows significant differences in EEG signals between trials where participants reported self-motion and those where they reported object-motion.
\autoref{fig:vection_forward} shows the difference between the runs where subjects reported \texttt{SV} versus the ones where subjects reported \texttt{WV} \& \texttt{NV}.
Since the signal displays similar patterns along the lateral axes, all three electrodes in each axe are averaged in order to reduce additive noise.
However, we also observed that vection events induce high signal variance, which widens our confidence intervals.

\begin{figure}[t]
    \begin{subfigure}{\columnwidth}
        \includegraphics[width=\linewidth]{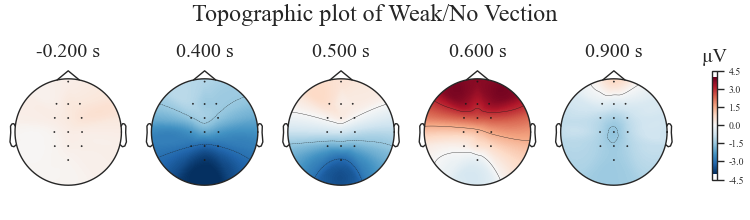}
    \end{subfigure}
    \begin{subfigure}{\columnwidth}
        \includegraphics[width=\linewidth]{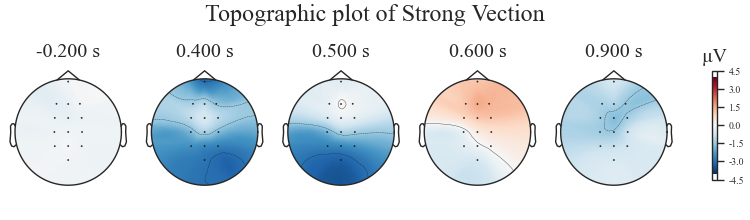}
    \end{subfigure}%
    \caption{Topographic map comparison of the average response for all subjects during $FA_1$ to No/Weak Vection (top) and Strong Vection (bottom). Strong vection displays a lingering parietal negativity until 600ms while Weak/No vection displays a stronger frontal positivity around 600ms}
    \label{fig:topo_vection}
\end{figure}

We plot the spacial differences using topographic maps between the vection conditions in \autoref{fig:topo_vection}.
The topographic maps highlight the stronger frontal positivity in the Weak/No Vection condition 600ms after acceleration onset, as well as the lingering parietal negativity in the Strong vection condition.

\begin{figure}[tb]
    \centering
    \includegraphics[width=\columnwidth]{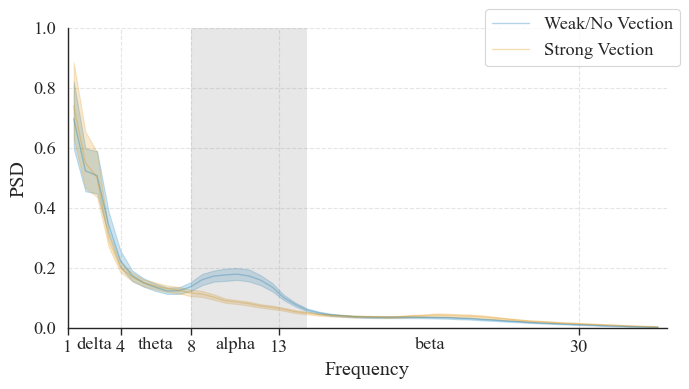}
    \caption{
    Power spectral density plot comparing \texttt{WV} \& \texttt{NV} (blue) to \texttt{SV} (orange) during $FA_1$ before filtering.
    The power of the frequencies during the acceleration event is shown, split between the vection conditions.
    The 95\% confidence interval is shown around the line of vection category.
    The shaded gray area corresponds to the time period where there was a statistically significant difference between the signals.
    $FA_1$ elicits a strong increase of the alpha rhythm, which is suppressed when the subject is experiencing self-motion.
    }
    \label{fig:psd}
\end{figure}

Additionally, \autoref{fig:psd} shows EEG rhythms associated with \texttt{SV} compared to \texttt{WV} \& \texttt{NV}.
Patterns consistent with the literature \cite{harquelModulationAlphaWaves2020, bertiNeuropsychologicalApproachesVisuallyInduced2020} are found, with alpha rhythm suppression during perceived self-motion.

Finally, we observe a link between reported vection and subject responses to the SSQ. 
Responses from each subject's pre-experiment questionnaire were subtracted from subsequent questionnaires to establish a baseline correction.
Most notably, there is a strong positive correlation as evidenced by the Pearson correlation between \texttt{SV} and the SSQ total score ($r=0.55$, $P<0.01$). 
More precisely, this correlation can be broken down in the SSQ subcategories: Oculomotor ($r=0.45$, $P<0.04$), Disorientation ($r=0.51$, $P<0.02$) and Nausea ($r=0.62$, $P<0.003$).
Thus, we observe that vection correlates the strongest with Nausea and Disorientation, which is consistent with the symptoms of cybersickness \cite{rebenitschReviewCybersicknessApplications2016}.

%% file: contents/discussion.tex
Our findings confirm previous results that identified distinct EEG markers associated with acceleration in VR.
Specifically, we observed significant signal differences between forward and backward motion in the FCz and Cz electrodes, consistent with earlier studies.
By analyzing EEG signals, we not only discerned the presence of acceleration but also identified the direction, with statistically distinct signals for forward versus backward motion.

For self-motion, we identified a notable neuromarker approximately 600 ms after the onset of acceleration, distinguishing strong vection experiences from weak or no vection experiences.
We notice a pattern during sudden changes in speed that varies depending on whether the subject experiences vection.
As illustrated in \autoref{fig:vection_forward}, a robust positive deviation 600ms after stimulus onset in the parietal region during strong vection gradually transitions into a negative deviation towards the frontal region, whereas the reverse pattern is observed for weak or no vection.

These results suggest that unpredictable visually induced motion in VR may trigger cognitive processes similar to those seen with real-world motion.
This is supported by the resemblance of our patterns to those observed in vestibular oddball paradigms by Nolan et al.\cite{nolanNeuralCorrelatesOddball2012}.
They investigate brain responses using a high-density electroencephalographic setup to an expected yet unpredictable event, also known as an oddball stimuli.
In this case, the stimulus was a movement of the chair the participant is sitting in.
The study found a robust P3 component, typical of oddball paradigms, indicating that vestibular changes in heading are processed similarly to oddball stimuli in other sensory modalities, with potential clinical relevance for assessing vestibular function.
In their paper, Figure 2 finds a pattern similar to the one we found in \autoref{fig:vection_forward}, we argue that the similarity in neural patterns hints to identical underlying cognitive processes.
Note that their paper induced vection using actual movements of the chair, thus inducing motion perception through visual and vestibular stimulations.
Consequently, we hypothesize that unpredictable visually induced motion in Virtual Reality triggers cognitive processes similar to those seen by unpredictable real-world motion.

A similar pattern was also found in the resolution of incongruity in the existing literature \cite{wangHumorDrawingsEvoked2017, duDifferentiationStagesJoke2013, tuNewAssociationEvaluation2014}.
Given that the sensation of vection originates from the brain interpreting the VR environment's acceleration as the participant's acceleration, we hypothesize that an incongruity arises due to sensory conflict.
This effect, we argue, is meaningfully detectable, as it pertains to the active resolution of this conflict.
Notably, this incongruity is not present when the participant does not experience vection, as they accurately perceive the environment as moving, rather than their body.
The absence of such signals in this context can be attributed to the absence of such incongruence.

This interpretation aligns to the sensory conflict theory of cybersickness.
Cybersickness is a major problem for widespread VR adoption, with 60--95\% of users affected and 6--42\% unable to finish their experiment \cite{nesbittCorrelatingReactionTime2017, casermanCybersicknessCurrentgenerationVirtual2021}.
Understanding the nature of cybersickness is crucial to improving the VR experience and increasing its adoption.
Sensory conflict theory is the most common theory of motion sickness and cybersickness \cite{palmisanoCybersicknessHeadMountedDisplays2020, davisSystematicReviewCybersickness2014}.
This theory posits the discrepancy between the visual, vestibular and proprioceptive senses as expectation and experiences cause cybersickness \cite{davisSystematicReviewCybersickness2014}.
This includes scenarios like perceiving self-motion in VR while remaining physically stationary. 
Our findings suggest that these neural responses may be indicative of the brain's active resolution of such sensory conflicts during vection experiences.
They provide valuable insights into the neurophysiological mechanisms underlying cybersickness.

We also found a correlation between vection and Simulator Sickness Questionnaire (SSQ) scores, particularly with Nausea and Disorientation scores.
Studies have highlighted the strong relationship between cybersickness and the Nausea and Disorientation scores \cite{kourtesisCybersicknessVirtualReality2023, nesbittCorrelatingReactionTime2017}. 
Therefore, the neuromarker associated with vection may serve as a precursor to cybersickness in VR participants.
As vection has often been associated with motion sickness \cite{nooijVectionMainContributor2017}, a better understanding of vection will allow for a deeper comprehension of motion sickness, its causes and mechanisms, and help VR designers create experiences that are more comfortable and less likely to cause sickness.
Moreover, detecting it in real time can help VR systems reduce motion sickness by adjusting content based on user feedback.
This research opens avenues for future exploration.
A promising direction lies in investigating the specific neuromarkers associated with backward vection.
Using a similar setup, studies could also explore if such signals can also be found in the case of sideways motion perception.
Additionally, the discovery of this neuromarker presents an exciting opportunity for the development of a classifier capable of determining whether a user is experiencing vection following an acceleration event.
Finally, this new paradigm can help study the neural mechanisms of vection in the brain.

%% file: contents/conclusion.tex
In this paper we studied if and how electroencephalography (EEG) could be used to detect vection in VR.
We conducted a VR experiment exposing participants to strong forward or backward acceleration while recording reported vection and EEG signals. 
Our results revealed substantial variability among individuals and a notable influence of the acceleration direction on reported vection. 
Replicating prior research, we observed a significant effect of vection on alpha power in EEG brain waves.
Moreover, the recorded EEG signals exhibited distinguishable patterns for both acceleration and direction of motion. 
Finally, we identified a new event related potential of vection occurring 600ms after forward acceleration.
These findings offer insights into vection's neural correlates, and pave the way to automatic techniques for measuring vection and self-motion sensations in VR using EEG recordings.